\documentstyle[festkoer,psfig]{vieweg}
\newcommand{\cugeo}{$\rm CuGeO_3$ }

\newcommand{\vanadat}{$\rm NaV_{2}O_{5}$ }

\newcommand{\cm}{~cm$\rm ^{-1}$ }

\newcommand{\ladder}{$\rm Sr_{14-x}Ca_xCu_{24}O_{41}$ }

\newcommand{\bfcugeo}{$\bf CuGeO_3$ }

\newcommand{\bfvanadat}{$\bf NaV_{2}O_{5}$ }
\newcommand{\bfladder}{$\bf Sr_{14-x}Ca_xCu_{24}O_{41}$ }

\author{\centerline{P. Lemmens, M. Fischer, M. Grove, P.H.M. v. Loosdrecht, G. Els,} \\
\centerline{E. Sherman$^{1}$, C. Pinettes$^{2}$, and G.
G\"untherodt}} \Kurzautor{P. Lemmens et al.}
\title{Quantum Spin Systems: From Spin Gaps to Pseudo Gaps}
\Kurztitel{From Spin Gaps to Pseudo Gaps}
\Adresse{2. Physikalisches Institut, RWTH
Aachen, Templergraben 55, D-52056 Aachen}

\begin{document}
\Titel

%%%%%%%%%%%%%%%%%%%%%%%%%%%%%%%%%%%%%%%%%%%%%%%%%%%%%%%%%%%%%%%%%%%%%%%%%%%
\begin{abstract}
Many low dimensional spin systems with a dimerized or ladder-like
antiferromagnetic exchange coupling have a gapped excitation
spectrum with magnetic bound states within the spin gap. For spin
ladders with an even number of legs the existence of spin gaps and
within the t-J model a tendency toward superconductivity with
d-wave symmetry is predicted. In the following we will
characterize the spin excitation spectra of different low
dimensional spin systems taking into account strong spin phonon
interaction ($\rm CuGeO_3$), charge ordering ($\rm NaV_2O_5$) and
doping on chains and ladders (\ladder). The spectroscopic
characterization of the model systems mentioned above has been
performed using magnetic inelastic light scattering originating
from a spin conserving exchange scattering mechanism. This is also
bound to yield more insight into the interrelation between these
spin gap excitations and the origin of the pseudo gap in high
temperature superconductors.
\end{abstract}

%For the latter system the behavior of the spin gap during the pressure- and
%doping-induced transition from the antiferromagnetic to the superconducting phase is
%of special interest to explain the pseudo gap effects observed in underdoped high
%temperature superconductors.
%%%%%%%%%%%%%%%%%%%%%%%%%%%%%%%%%%%%%%%%%%%%%%%%%%%%%%%%%%%%%%%%%%%%%%%%%%%%%

%%%%%%%%%%%%%%%%%%%%%%%%%%%%%%%%%%%%%%%%%%%%%%%%%%%%%%%%%%%%%%%%%%%%%%%%%%%%%
\section{Introduction}
%%%%%%%%%%%%%%%%%%%%%%%%%%%%%%%%%%%%%%%%%%%%%%%%%%%%%%%%%%%%%%%%%%%%%%%%%%%%%
\setcounter{footnote}{1} \footnotetext{and Moscow Inst. of Physics
and Technology, 141700 Dolgoprudny, Russia} \stepcounter{footnote}
\footnotetext{and LPTM, Univ. de Cergy, 2 Av. A. Chauvin, 95302
Cergy-Pontoise Cedex, France}

There is a general consensus that part of the unusual physics of
doped two-dimensional spin systems, i.e. the observation of pseudo
gaps and high temperature superconductivity, can be mapped onto
one dimension. As the pseudo gaps are evident not only in
transport and thermodynamic measurements but also in NMR
spectroscopy they certainly involve spin degrees of freedom. It
was predicted that the binding of mobile holes in spin ladders can
lead either to a superconducting or a charge-ordered ground state.
The observation of superconductivity in the spin ladder/chain
compound \ladder and the discussion of a phase separation into 1D
spin and charge stripes in high temperature superconductors (HTSC)
and related compounds encouraged this assumption
\cite{sigrist94,levi96,dagotto96b,tranquada96}. However, since for
\ladder there is some evidence of a crossover toward a
two-dimensional system \cite{nagata98} and a possible vanishing of
the spin gap under pressure \cite{mayaffre98} it is not clear
whether two-leg or the recently studied three-leg ladders provide
useful analogs to HTSC \cite{rice97}. Therefore, an investigation
of the excitation spectrum of low dimensional spin systems, in
particular in compounds with a spin gap is important and may shed
some light on the similarities and differences between both
classes of materials.

%%%%%%%%%%%%%%%%%%%%%%%%%%%%%%%%%%%%%%%%%%%%%%%%%%%%%%%%%%%%%%%%%%%%%%%%%%%%%%%%%%%
\section{Structural Elements of Low Dimensional Spin Systems}
%%%%%%%%%%%%%%%%%%%%%%%%%%%%%%%%%%%%%%%%%%%%%%%%%%%%%%%%%%%%%%%%%%%%%%%%%%%%%%%%%%%
\begin{figure}[t]
\centerline{\psfig{file=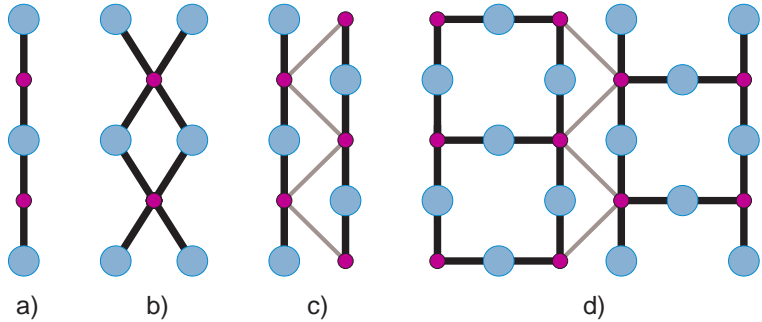,width=10cm}} \caption[Overview
of 3d ion-oxygen configurations]{\label{uebers} Overview of 3d
ion-oxygen configurations realized in low dimensional transition
metal compounds: a) a simple 3d ion-O-chain, b) a non linear 3d
ion-O$_2$-chain with reduced exchange, c) a frustrated double
chain (zigzag chain), and d) two ladders with a frustrated weak
coupling. The thick (thin) lines mark strong (weak) exchange
coupling paths. The small circles denote the positions of the
transition metal ions, e.g. Cu$^{2+}$ with s=1/2. The large
circles denote O$^{2-}$ \cite{lemmens-hab}.}
\end{figure}

In the systems discussed here the low energy excitations are
mainly due to the spin degrees of freedom. The magnetic properties
may often be described by the Heisenberg exchange spin
Hamiltonian. If, in addition, the exchange is restricted to low
dimensions then chains, spin ladders, and further systems with a
more complex exchange pattern are realized.

Two building principles are used to reduce the superexchange of a
3d ion-oxygen configuration to less than three dimensions. These
are on the one hand an enlarged distance or missing bridging
oxygen between two 3d ion-sites or on the other hand a
superexchange path with an angle close to 90$^{\circ}$. Due to the
Kanamori-Goodenough rule (vanishing superexchange via
perpendicular oxygen O2p-orbitals) a non collinear exchange path
leads to a magnetic insulation of, e.g. neighboring CuO chains. In
this way compounds representing chains, zigzag double chains or
ladders with different numbers of legs are realized.
Fig.~\ref{uebers} shows a comparison of several possible 3d
ion-oxygen configurations. Compounds that incorporate these
structural elements exhibit a number of unusual properties which
are related to strong quantum fluctuations.

%%%%%%%%%%%%%%%%%%%%%%%%%%%%%%%%%%%%%%%%%%%%%%%%%%%%%%%%%%%%%%%%%%%%%%%%%%%%%
\section{Excitation Spectrum and Phase Diagram}
%%%%%%%%%%%%%%%%%%%%%%%%%%%%%%%%%%%%%%%%%%%%%%%%%%%%%%%%%%%%%%%%%%%%%%%%%%%%%
The excitation spectrum of a one-dimensional spin system (spin
chain) with nearest neighbor exchange coupling is characterized by
a degeneracy of the singlet ground state with triplet excitations
in the thermodynamic limit \cite{auerbach}. Assuming negligible
spin anisotropies the ground state is not magnetically ordered
even for T=0 and there are gapless excitations. The spin-spin
correlations are algebraically decaying. The elementary
excitations in such a system are therefore described as massless
asymptotically free pairs of domain wall-like solitons or s=1/2
spinons. A quantum phase transition from this gapless critical
state into a gapped spin liquid state may be induced by a
dimerization, i.e. an alternation of the coupling constants
between nearest neighbors, or by a sufficient frustration due to
competing next nearest neighbor antiferromagnetic exchange. This
gapped state is characterized by extremely short ranged spin-spin
correlations and may be described as an arrangement of weakly
interacting spin dimers \cite{bray83}.

A simple representative of the quantum disordered state is the
two-leg spin ladder with a larger exchange coupling along the
rungs than along the legs of the ladder \cite{dagotto92}. The
singlet ground state is composed of spin dimers on the rungs. An
excitation in the picture of strong dimerization corresponds to
breaking one dimer leading to a singlet-triplet excitation $\rm
\Delta_{01}$. Studies on three-, four- or five-leg ladders led to
the conjecture that ladders with an even number of legs have a
spin gap while odd-leg ladders are gapless
\cite{levi96,dagotto96b}. A family of compounds that may represent
these systems are the Sr cuprates, e.g. the two-leg ladder
compound $\rm SrCu_{2}O_3$ \cite{azuma94} and the system \ladder
that is composed of a chain and a ladder subcell and moreover
shows superconductivity under pressure \cite{uehara97}.

In the limit of an infinite number of coupled chains a
two-dimensional Heisenberg system is obtained and the spin gap
vanishes. This limit has also been used to study the
two-dimensional high temperature superconductors. Within this
framework, also weakly doped two- and three-leg ladder were
theoretically investigated \cite{sigrist94,rice97,rice93}.

%In 2D a spin dimer ground state with a gapped excitation spectrum is also realized in
%the strongly frustrated system $\rm SrCu_2(BO_3)_2$ \cite{miyahara98}.

%%%%%%%%%%%%%%%%%%%%%%%%%%%%%%%%%%%%%%%%%%%%%%%%%%%%%%%%%%%%%%%%%%%%%%%%%%%%%%%%%
\section{Magnetic Bound States in \bfcugeo and \bfvanadat}
%%%%%%%%%%%%%%%%%%%%%%%%%%%%%%%%%%%%%%%%%%%%%%%%%%%%%%%%%%%%%%%%%%%%%%%%%%%%%%%%%
\begin{figure}
\centerline{\psfig{file=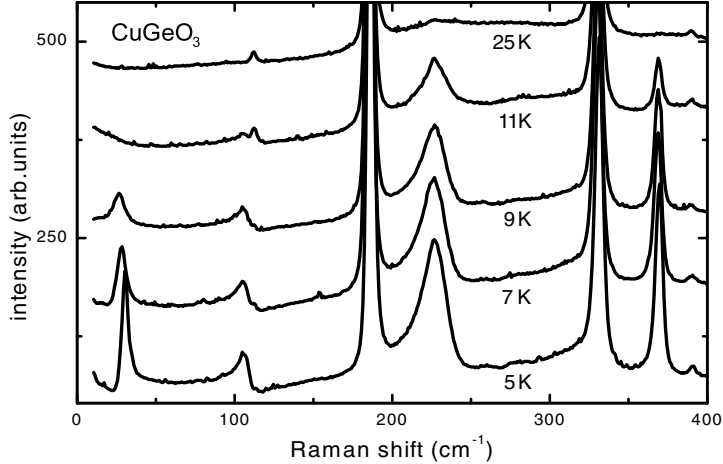,width=9.5cm}}
\caption[Raman spectra of $\rm CuGeO_3$] {\label{cugeo1}
Intrachain (cc) polarized Raman light scattering spectra of $\rm
CuGeO_3$ for temperatures above and below $\rm T_{SP}$=14~K
\cite{lemmens97}. For T$>$$\rm T_{SP}$ a broad continuum is
observed around 250~cm$^{-1}$ whereas for T$<$$\rm T_{SP}$ several
dimerization-induced modes are observed. }
\end{figure}

%\begin{figure}
%\centerline{\psfig{file=bilder/cugeosub1b.eps,width=11cm}} \caption[Intensities of the
%dimerization-induced modes in $\rm CuGeO_3$] {\label{cugeo2}Renormalized scattering
%intensities of the dimerization-induced modes in $\rm CuGeO_3$. The integrated
%intensity of the continuum is denoted by open squares. The phonon modes at 369~cm$\rm
%^{-1}$, 224~cm$\rm ^{-1}$ and the singlet bound state at 30~cm$\rm ^{-1}$ are given by
%closed squares, closed triangles and open circles respectively \cite{lemmens97}.}
%\end{figure}

A salient feature of low dimensional quantum spin systems with a
gapped excitation spectrum is the existence of magnetic bound
states, i.e. triplet excitations that are confined to bound
singlet or triplet states
\cite{uhrig96,bouz98,affleck98,sushkov98}. These states are
characterized by a well-defined excitation with an energy reduced
with respect to the energy of a two-particle continuum of "free"
triplet excitations. In the case of a spin chain the binding
energy originates from frustration and/or interchain interaction.
In general, these states may therefore be used to study the
triplet-triplet interaction, the coupling parameters and the phase
diagram of the system.

%If interchain or magnetoelastic interactions are dominant these states are best
%characterized as soliton-antisoliton bound states
%\cite{affleck98,augier98,sorensen98,augier99}.

Magnetic bound states of singlet character may be investigated
using light scattering experiments. The light scattering process
involved results from a spin-conserving exchange mechanism
\cite{muthu96,gerd97}. For these investigations spin-Peierls
compounds are very promising as they show a transition from a
homogeneous to a dimerized phase for temperatures below the
spin-Peierls temperature $\rm T_{SP}$. Therefore, excitations of
these systems may be characterized due to their behavior in
dependence on temperature, i.e. as function of the dimerization of
the spin system. Magnetic bound states have been identified in
light scattering experiments on $\rm CuGeO_3$ as a single
\cite{gerd97} and on \vanadat as multiple singlet states
\cite{lemmens98}.

Raman spectra of \cugeo shown in Fig.~\ref{cugeo1} show for
T$<$$\rm T_{SP}$=14~K additional dimerization-induced modes which
are zone-folded phonons with the exception of one mode at
30~cm$^{-1}$. The Fano-lineshape of these modes at 104\cm and
224\cm is caused by spin-phonon coupling. The excitation at $\rm
30~cm^{-1}$ is identified as a singlet bound state. Its energy
$\rm \Delta_{00}$=30\cm$\rm \approx\sqrt{3}\Delta_{01}$, with $\rm
\Delta_{01}$=16.8\cm the singlet-triplet gap and the quasi-linear
increase of its intensity with decreasing temperature support this
interpretation \cite{lemmens97,gerd97}.
%The temperature dependence of the
%intensity of these modes is given in Fig.~\ref{cugeo2}.

\begin{figure}
\centerline{\psfig{file=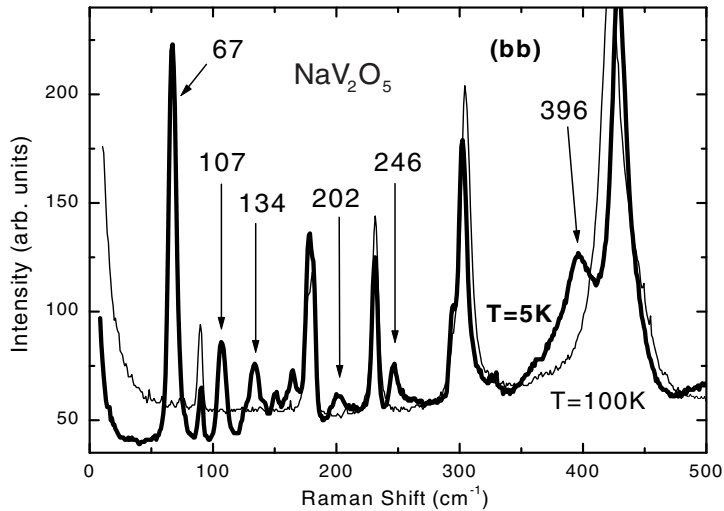,width=9.5cm}}
\caption[Raman spectra of $\rm NaV_{2}O_{5}$ at 100 and 5~K]{Raman
light scattering spectra of $\rm NaV_{2}O_{5}$ at 100 and 5~K with
incident and scattered light parallel to the (bb) ladder
direction. The additional modes in the low temperature phase are
marked by arrows.} \label{navo}
\end{figure}

Corresponding experiments on the compound \vanadat with T$\rm
_{SP}$=34~K \cite{isobe96} given in Fig.~\ref{navo}, show more
transition-induced modes. Using the criteria discussed above,
three modes at 67, 107 and 134\cm are candidates for singlet bound
states. In addition there is a decrease of the background
scattering intensity for frequencies $\Delta\omega$$<$120\cm which
is indicative of 2$\Delta\rm _{01}$ in agreement with magnetic
susceptibility data \cite{weiden97c}. This compound differs from
the spin chain system \cugeo in the sense that it represents a
quarter-filled spin ladder that only for T$>$ T$_{\rm SP}$ may be
mapped on a spin chain \cite{smolinski98}. Furthermore, there is
strong evidence that the transition at T$\rm _{SP}$ is not a
spin-Peierls transition but an electronically driven dimerization
connected with a charge ordering of the s=1/2 V$\rm ^{4+}$ and
V$^{5+}$ on the rungs of the ladders
\cite{thalmeier98,mostovoy98,seo98}.

%%%%%%%%%%%%%%%%%%%%%%%%%%%%%%%%%%%%%%%%%%%%%%%%%%%%%%%%%%%%%%%%%%%%%%%%%%%%%%%%%%
\section{The Doped Chain/Ladder System \bfladder}
%%%%%%%%%%%%%%%%%%%%%%%%%%%%%%%%%%%%%%%%%%%%%%%%%%%%%%%%%%%%%%%%%%%%%%%%%%%%%%%%%%
\begin{figure}
\centerline{\psfig{file=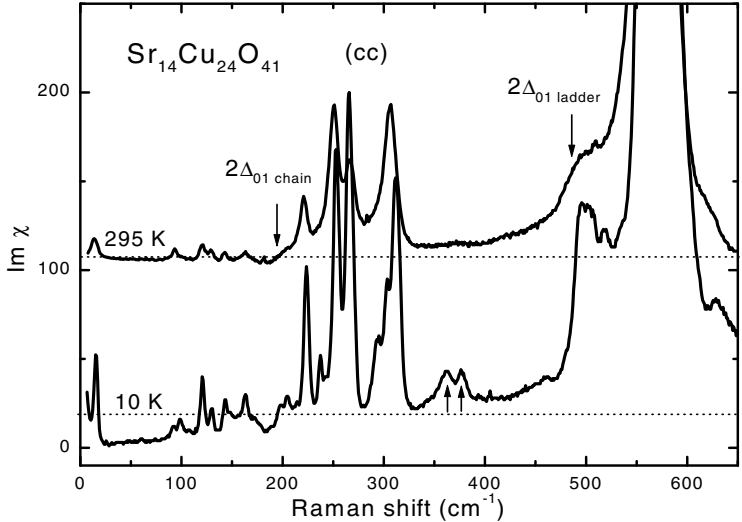,width=9.8cm}}
\caption[Raman light scattering spectra of \ladder]{Raman light
scattering spectra of \ladder with x=0 in intraladder (cc)
polarization (the curves have been given an offset for clarity).
The doubled gaps of the chain and the ladder system as well as
additional modes are marked by arrows. To emphasize the small
redistribution of spectral weight the background of the scattering
intensity at high frequencies is indicated by a dotted line
\cite{grove-o41}.} \label{ladder}
\end{figure}

\begin{figure}
\centerline{\psfig{file=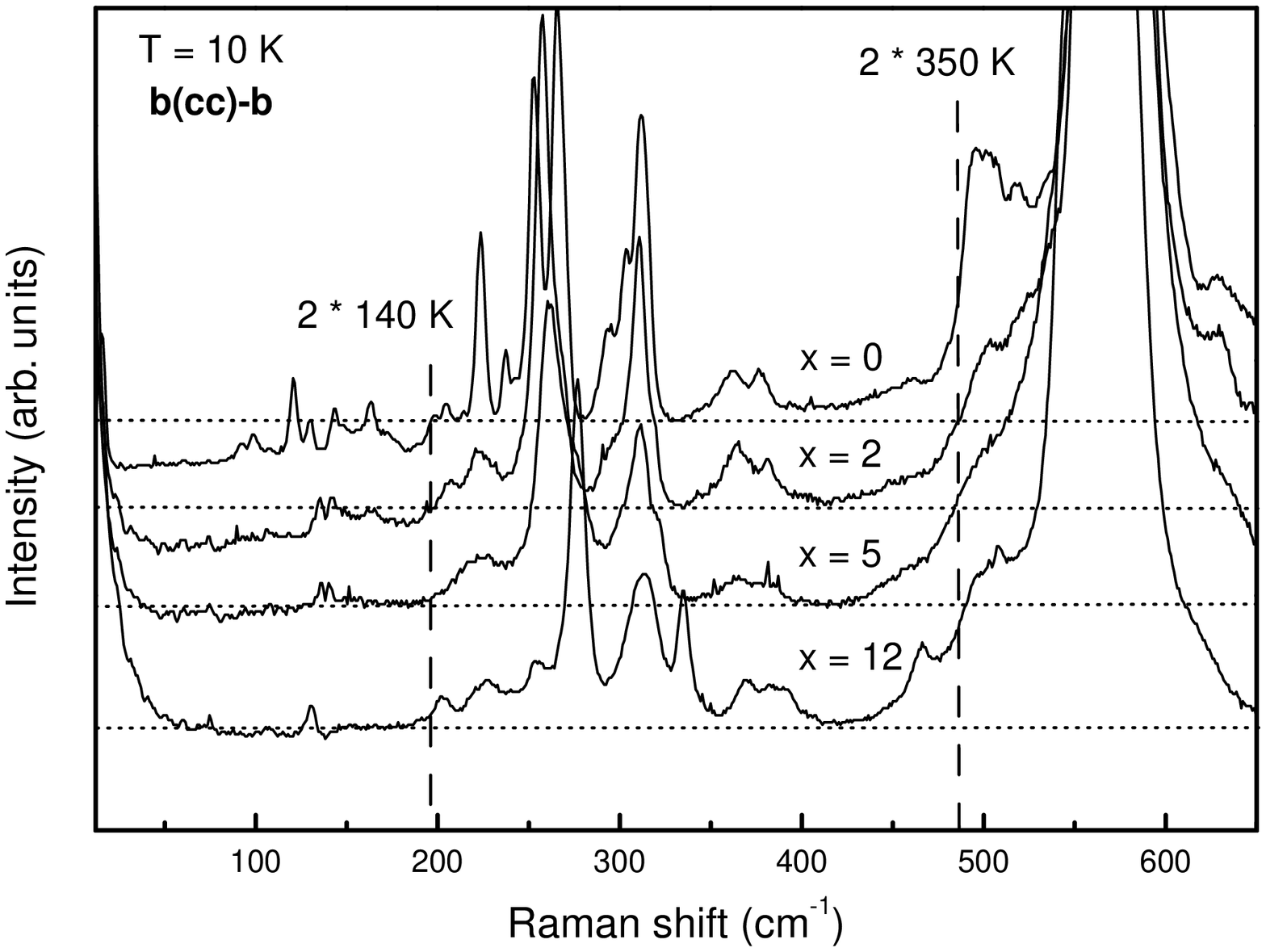,width=10cm}} \caption[Raman
light scattering spectra of \ladder]{Raman light scattering
spectra of \ladder in dependence on Ca substitution with x=0, 2,
5, 12 (intraladder (cc) polarization) \cite{grove-o41}.}
\label{ladder2}
\end{figure}

In the compound \ladder, which incorporates both $\rm CuO_2$
chains and $\rm Cu_{2}O_{3}$ ladders, a substitution of Sr by the
isovalent Ca together with applied pressure leads first to a
transfer of holes from the chains to the ladders followed by a
delocalization of the holes \cite{osafune97}. Superconductivity is
observed for pressures around 3GPa with a maximum transition
temperature of T$\rm _c$=12~K \cite{uehara97}. Ca-substitution and
applying pressure reduces the b and c axis parameters leading to
strong changes of the electronic properties, e.g., a reduction of
the anisotropy in the resistivity. In samples with x=11.5 the
anisotropy of the resistivity $\rho_a$/$\rho_c$ at T=50~K
decreases from 80 (P=0) to 10 (P=4.5~GPa), i.e. it shifts towards
a more two-dimensional behavior \cite{nagata98}.

For x=0 a singlet-triplet gap in the chains, $\Delta\rm
_{01~chain}$=140~K or 125~K, has been determined using magnetic
susceptibility \cite{carter96} and NMR experiments
\cite{takigawa98}, while a gap in the ladder of $\Delta\rm
_{01~ladder}$=375~K in neutron scattering experiments
\cite{katano99} or 550~K in NMR \cite{magishi98} has been
observed. For x$\neq$0 the gap in the chain system rapidly
disappears. However, the effect on the gap in the ladder system is
unclear. While in NMR experiments a strong decrease of the gap
with substitution from $\Delta\rm _{01~ladder}$=550~K (x=0) to
270~K (x=11.5) has been observed \cite{magishi98}, the
corresponding neutron experiments show no change at all
\cite{katano99}. In optical conductivity measurements inspired by
similar results in HTSC the opening of a "pseudo gap" is claimed
\cite{osafune99}. Finally, with applied pressure NMR experiments
indicate a change of the gap in the ladder to a "pseudo spin gap"
\cite{mayaffre98}. Although the coexistence of this gap with
superconductivity would be a very important piece of evidence,
these results could up to now neither be proved nor disproved by
other methods.

Concerning the origin of the smaller gap in the chains a
dimerization and charge ordering is discussed. Indeed,
superstructure peaks that increase in intensity for temperatures
below 50~K are observed in X-ray scattering on samples with x=0
\cite{cox98}. Surprisingly, the corresponding dimers are formed in
the chains between the Cu spins that are separated by 2 times the
distance between the nearest neighbor Cu ions. The distance
between two neighboring dimers is 4 times the distance of nearest
neighbor Cu ions. Therefore, the dimerization corresponds to
ordered Zhang-Rice singlets on the chains. The importance of these
singlet states is also discussed for the 2D HTSC \cite{sigrist94}.
In NMR experiments on \ladder the existence of both Cu$\rm ^{2+}$
and Cu$\rm ^{3+}$ in the chains has been verified
\cite{takigawa98}.

In Raman scattering experiments with light polarization parallel
to the ladder direction (bb) both gaps are identified as a
renormalization of the scattering intensity to lower frequency at
2$\Delta\rm _{01~chain}$=280~K and 2$\Delta\rm
_{01~ladder}$=700~K. These values are close to the frequencies
found in the above discussed neutron experiments (see
Fig.~\ref{ladder} and phonon spectra in Ref.~\cite{abrashev97}),
and differ substantially from the NMR results.

The signatures of both chain and ladder gaps weaken and broaden
with increasing temperature till they disappear for temperatures
above 100 and 350~K for the chain and ladder, respectively.
Furthermore, additional modes are observed at low temperatures at
360 and 375~cm$\rm ^{-1}$. Although these modes may be phonons, it
is interesting to note that their energies correspond to
1.48$\Delta\rm _{01~ladder}$ and 1.54$\Delta\rm _{01~ladder}$,
respectively, making them candidates for singlet bound states of
the ladder. In Fig.~\ref{ladder2} Raman spectra of \ladder with
different x=0, 2, 5 and 12 are compared. Strong changes of the
phonon lines in the frequency range
120\cm$<$$\Delta\omega$$<$350\cm are evident that may be related
to a change of the commensurability of the chain and ladder
subcells. In addition, the gap of the chain subsystem is
suppressed with increasing Ca substitution. In contrast to these
effects, the signature of the gap in the ladder subsystem is only
broadened but not shifted in frequency. This supports the
negligible substitution dependence of $\Delta\rm _{01~ladder}$
observed in neutron scattering. The additional modes that are
tentatively attributed to bound states are also not influenced by
Ca substitution.

%%%%%%%%%%%%%%%%%%%%%%%%%%%%%%%%%%%%%%%%%%%%%%%%%%%%%%%%%%%%%%%%%%%%%%%
\section{Conclusion}
%%%%%%%%%%%%%%%%%%%%%%%%%%%%%%%%%%%%%%%%%%%%%%%%%%%%%%%%%%%%%%%%%%%%%%%
The low energy excitation spectrum of low dimensional spin systems
has been under intense investigation during the last years. Both
CuGeO$_{3}$ and NaV$_{2}$O$_{5}$ can be considered as model
compounds as a spin gap opens below a phase transition temperature
T$_{\rm SP}$. In inelastic light scattering experiments this spin
gap is evidenced by a renormalization of the background intensity
below 2$\Delta_{01}$. Furthermore, well defined singlet bound
states consisting of two triplet excitations are found. As their
multiplicity and binding energy crucially depend on system
parameters their analysis gives a wealth of information on the
principal magnetic interactions in the system. These bound states
are the magnetic analog of exciton states in semiconductors.

It has been argued that the pseudo gap in HTSC can be understood
in terms of a spin gap. In this context, the investigation of
\ladder should be very useful, as this substance consists of both
ladders and dimerized chains and becomes superconducting. It
therefore can be understood as a link between the low dimensional
spin gap systems and HTSC. Inelastic light scattering on \ladder
samples with x=0 shows, in close analogy to $\rm CuGeO_{3}$ and
$\rm NaV_{2}O_{5}$, a drop in intensity for frequencies below
2$\rm \Delta_{01}$. Possibly, magnetic bound states for the ladder
emerge as well. For x~$\neq$~0 the gap in the chains vanishes in
agreement with results of other methods. On the other hand, the
gap of the ladder persists even for x=12, the doping concentration
for which superconductivity occurs under applied pressure. It will
be of particular interest to follow the evolution of the spin gap
approaching the superconducting phase. Therefore, measurements
under hydrostatic pressure are highly desirable and under
preparation. A comparison of NMR, neutron and Raman scattering
results shows that the first method does not sample the same
physical quantity as the other two. This problem is not fully
understood. The question how this spin gap and the pseudo gap as
observed in HTSC are related could be addressed by these
investigations. It is questionable whether the spin gap and the
pseudo gap can be directly identified in \ladder as proposed in
Ref.~\cite{osafune99} as the energy scales of the superconducting
gap and the spin gap are different by more than an order of
magnitude. Nevertheless, the study of these low dimensional
quantum spin systems is of fundamental importance for the
understanding of collective quantum phenomena in strongly
correlated electron systems, such as magnetism and
superconductivity.

%%%%%%%%%%%%%%%%%%%%%%%%%%%%%%%%%%%%%%%%%%%%%%%%%%%%%%%%%%%%%%%%%%%%%%%%%%%%%
{\bf Acknowledgement:} Single crystalline samples were kindly
provided by M. Weiden, E. Morre, C. Geibel, and F. Steglich
(MPI-CPfS, Dresden), U. Ammerahl, G. Dhalenne, and A. Revcolevschi
(Univ. Paris-Sud), J. Akimitsu (Aoyama-Gakuin Univ., Tokyo).
Further support by the DFG under SFB341 and the BMBF under Fkz
13N6586/8 and 13N7329/3, and by the INTAS Project 96-410 are
kindly acknowledged.

% H. Kageyama, and Y. Ueda (ISSP Tokyo), H. Tanaka (Tokyo Inst. Techn.),
% B.C. Sales (Oak Ridge Nat. Lab.), F. B\"ullesfeld, and
% W. Assmus (Inst. f. Physik, Univ. Frankfurt)

%%%%%%%%%%%%%%%%%%%%%%%%%%%%%%%%%%%%%%%%%%%%%%%%%%%%%%%%%%%%%%%%%%%%%%%%%%%%%%

%%%%%%%%%%%%%%%%%%%%%%%%%%%%%%%%%%%%%%%%%%%%%%%%%%%%%%%%%%%%%%%%%%%%%%%%%%%%%%%
%\bibliography{lit}
%%\bibliographystyle{plain}
%\bibliographystyle{peter}
%% print out full name list

%\begin{thebibliography}{99}
%\bibitem{kopka} H. Kopka, {\em LATEX - Eine Einf\"uhrung.} (Addison
%Wesley, Reading(MA) 199y)
%\bibitem{lamport}L. Lamport, {\em LATEX---A Document Preparation
%System.} (Addison Wesley, Reading(MA) 1985)
%\bibitem{journals} F. Author, S. Author, Journ. Name B{\bf vol},
%page (year).
%\end{thebibliography}

\end{document}